\documentclass[12pt,preprint]{aastex}

\usepackage{graphicx}

\slugcomment{Astronomical Journal, in press 2013-Mar-08}

\shorttitle{Hot Bodies}
\shortauthors{Jewitt}

\begin{document}

\title{Properties of Near-Sun Asteroids\footnote
{The data presented herein were obtained at the W.M. Keck Observatory, which is operated as a scientific partnership amongst the California Institute of Technology, the University of California and the National Aeronautics and Space Administration. The Observatory was made possible by the generous financial support of the W.M. Keck Foundation. 
}   
}

\author{David Jewitt$^2$}
\affil{$^2$Department of Earth and Space Sciences and Department of Physics and Astronomy,
University of California at Los Angeles, 
595 Charles Young Drive East, \\
Los Angeles, CA 90095-1567\\
}

\email{jewitt@ucla.edu}

\begin{abstract}
Asteroids near the Sun can attain equilibrium temperatures sufficient to induce surface modification from thermal fracture, desiccation and decomposition of hydrated silicates.  We present optical observations of nine asteroids with perihelia $<$0.25 AU (sub-solar temperatures $\ge$800 K) taken to search for evidence of thermal modification.  We find that the broadband colors of these objects are diverse but statistically indistinguishable from those of planet-crossing asteroids having  perihelia near 1 AU.  Furthermore, images of these bodies taken away from perihelion show no evidence for on-going mass-loss (model-dependent limits $\lesssim$ 1 kg s$^{-1}$) that might result from thermal disintegration of the surface.  We conclude that, while thermal modification may be an important process in the decay of near-Sun asteroids and in the production of debris, our new data provide no evidence for it.  
\end{abstract}

\keywords{minor planets, asteroids: general}

\section{Introduction}
Some asteroids from the main belt are scattered into eccentric orbits having perihelia at small heliocentric distances.  These objects are dynamically short-lived because of encounters with the terrestrial planets, whose orbits they cross.  In extreme cases, the equilibrium surface temperatures reached at perihelion can surpass the temperature needed to cause desiccation of hydrated silicates, possibly leading to modification of the spectral reflectance properties of the bodies (Akai 1992, Hiroi et al.~1996, Cloutis et al.~2012).  Thermal fracture of exposed rocks may also occur.  Both desiccation cracking and thermal fracture are potentially capable of generating dust (Jewitt and Li 2010, Jewitt 2012).   

Evidence for this has been reported in asteroid (3200) Phaethon which, with a perihelion distance $q$ = 0.14 AU, can reach subsolar temperatures up to $\sim$1000 K (Ohtsuka et al.~2009, Jewitt and Li 2010). Phaethon has exhibited anomalous brightening on two occasions when at perihelion (Jewitt and Li 2010, Li and Jewitt 2013).  The brightening is inconsistent with the phase function of any known macroscopic body, and is too large to be attributed to thermal emission or to fluorescence excited by solar wind impact or by solar ultraviolet photons.  Phaethon is also too hot for water ice to play any role in the observed activity.  The remaining viable hypothesis for the cause of the anomalous perihelion brightening is mass loss in the form dust, leading to a transient, increased cross-section for the scattering of sunlight.    Phaethon is the parent of the Geminid meteor stream and a member of the so-called Phaethon-Geminid-Complex (Kasuga 2009, Ohtsuka et al.~2009).  Dust production at perihelion may thus be seen as contributing to the production of Geminids, although the fraction of the Geminid mass that can be supplied in this way is highly uncertain.  Laboratory experiments  show that the optical properties of meteorites can be altered by heating, while the diversity of asteroid reflection spectra has long been interpreted in terms of thermal metamorphism (Hiroi et al.~1966).

Motivated in part by the results from Phaethon, we have obtained observations of other near-Sun asteroids (NSAs).   We ask two questions related to these objects.  First, do the NSAs share any common properties, for example the optical colors, that might be related to the high temperatures experienced at perihelion?  Second, is there evidence of on-going mass loss from NSAs in high resolution, high sensitivity optical data?  With few exceptions, the new data are the first reported physical measurements of the NSAs, either individually or as a group.


\section{Observations} 

We used the 10-meter diameter Keck I telescope located atop Mauna Kea, Hawaii and the Low Resolution Imaging Spectrometer (LRIS) camera (Oke et al.~1995) to image these objects. The LRIS camera has two channels housing red and blue optimized charge-coupled devices separated by a dichroic filter (we used the ``460'' dichroic, which has 50\% transmission at 4875\AA).  On the blue side we used a broadband B filter (center wavelength $\lambda_c$ = 4369\AA, full width at half maximum (FWHM) $\Delta \lambda$ = 880\AA) and on the red side an R filter ($\lambda_c$ = 6417\AA, $\Delta \lambda$ = 1185\AA).  All observations used the facility atmospheric dispersion compensator to correct for differential refraction, and the telescope was tracked non-sidereally while autoguiding on fixed stars (except for 2002 AJ129, where the autoguider failed).   The image scale on both cameras was 0.135\arcsec~pixel$^{-1}$ and the useful field of view approximately 320\arcsec$\times$440\arcsec.  Atmospheric seeing ranged from  $\sim$0.7 to 1.3\arcsec~FWHM and the sky above Mauna Kea was photometric.  

The data were reduced by subtracting a bias (zero exposure) image and then dividing by a flat field image constructed from integrations taken on a diffusely illuminated spot on the inside of the Keck dome.  The NSAs were identified in the flattened images from their positions and their distinctive sky-plane motions.  Photometry was obtained using circular projected apertures tailored to the individual nightly observing conditions.  In seeing of 1\arcsec~FWHM and less, typical aperture radii were 1.5\arcsec~to 2.0\arcsec, with sky subtraction obtained from a contiguous annulus having outer radius 3.3\arcsec~to 6.6\arcsec.  These selections ensure that $>$90\% of the light from the image was captured within the photometry aperture.  In poorer seeing we used appropriately larger apertures.  Photometric calibration was secured from observations of standard stars from Landolt (1992), always using the same apertures as employed for the target asteroids.

The eccentric orbits and small sizes of the NSAs mean that they are typically either located at inconveniently small angles from the Sun or are uncomfortably faint, when far from the Sun.  We observed them opportunistically using observing time allocated to other projects, when their apparent magnitudes were bright enough and their solar elongations large enough to make them easily measurable with a minimal investment of observing time.  In some cases, we observed in conditions of sub-standard seeing, when our main projects (on much fainter outer solar system targets) were impossible.  Consequently, the average seeing for the present measurements was below values normally associated with Mauna Kea.  However, in all cases we secured photometry of adequate signal-to-noise ratio.

The orbital parameters of the target objects are listed in Table (\ref{orbits}).  The Table also lists the isothermal, spherical blackbody temperature at perihelion distance, $q$, computed from $T_{BB}(q) = \left(F_{\odot}/(4 \sigma q^2 )\right)^{1/4}$ and the sub-solar temperature $T_{SS}(q) = 2^{1/2}T_{BB}(q)$.  Here, $F_{\odot}$ = 1360 W m$^{-2}$ is the solar constant, $\sigma$ = 5.67$\times$10$^{-8}$ W m$^{-2}$ K$^{-4}$ is the Stefan-Boltzmann radiation constant and $q$ is expressed in AU.  The geometrical circumstances of observation are given for each object in Table (\ref{geometry}).  

\begin{figure}
\epsscale{1.00}
\begin{center}
\plotone{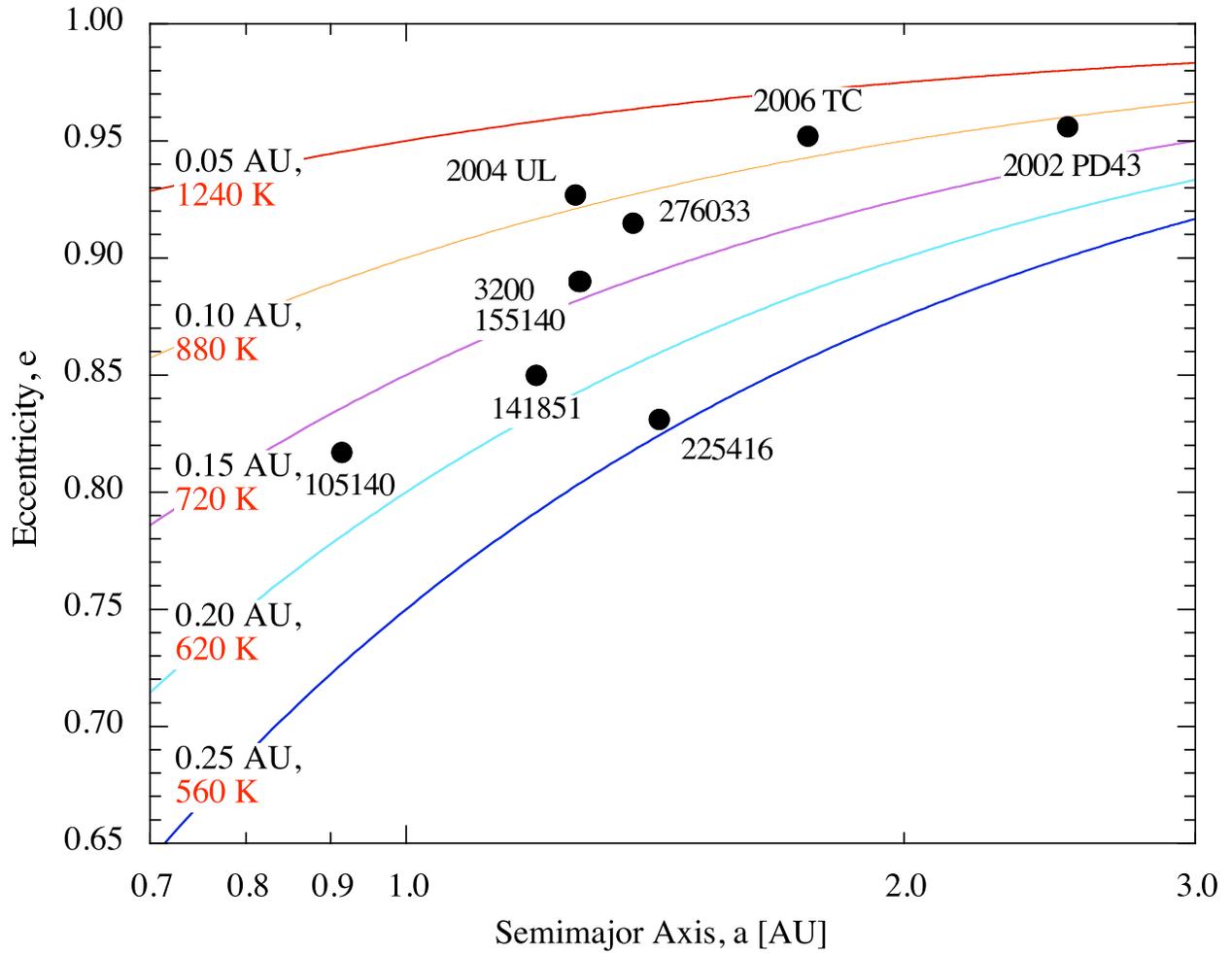}
\caption{Orbital semimajor axis vs.~eccentricity for the objects of this study.  Curves mark loci of constant perihelion distance, labeled in black.  Red labels on the curves show the corresponding isothermal, spherical blackbody temperature for each perihelion distance.  \label{ae}
} 
\end{center} 
\end{figure}

Figure (\ref{ae}) shows the semimajor axis vs.~orbital eccentricity plane, with lines of constant perihelion distance, $q$, marked.  Objects in our sample span the range 0.09 $\le q \le$ 0.24 AU, corresponding to 570 $\le T_{BB} \le$ 930 K and 800 $\le T_{SS} \le$ 1310 K.  This is the range of temperatures over which, for example, hydrated minerals decompose and which is therefore of interest in this study.

\section{Results}

\subsection{Photometry and Colors} 

The apparent magnitudes and colors are summarized in Table (\ref{photometry}), with uncertainties estimated from the scatter of repeated measurements.  One object, 2000 NL10, was independently observed by Dandy et al.~(2003).  Their determinations of B-V = 0.86$\pm$0.02 and V-R = 0.42$\pm$0.02 are close to ours (B-V = 0.83$\pm$0.05 and V-R = 0.40$\pm$0.04, Table \ref{photometry}).  

Next, the apparent magnitudes were converted to absolute magnitudes (i.e.~scaled to unit heliocentric, geocentric distances and at 0\degr~phase angle) using

\begin{equation}
H_R = m_R - 5\log_{10}\left(R_{au} \Delta_{au} \right) + 2.5\log_{10}(\Phi(\alpha))
\label{inverse}
\end{equation}

\noindent in which $R_{au}$ and $\Delta_{au}$ are the heliocentric and geocentric distances, respectively, both expressed in AU, and $\Phi(\alpha)$ is the ratio of the brightness at phase angle $\alpha$ to that at phase angle 0\degr.   We employed the HG formalism (Bowell et al.~1989) with scattering parameter $g$ = 0.25, as appropriate for an S-type asteroid.  To show the importance of the phase angle correction, we also list in Table (\ref{photometry}) the magnitude from Equation (\ref{inverse}) computed assuming 2.5$\log_{10}(\Phi(\alpha)$ = 0.  Evidently, the phase correction is substantial as a result of the large phase angles of observation, and for many objects exceeds 1 mag.  Uncertainties in $H_R$ are several $\times$0.1 mag., but are not known in detail in the absence of additional information about the phase functions of the NSAs.  We note that $H_R$ = 17.5, with assumed red geometric albedo $p_R$ = 0.15,  corresponds roughly to an asteroid diameter of 1 km.  Given this, the objects in Table (\ref{photometry}) have diameters in the 0.5 km to 5 km range.  Accurate diameters cannot be determined from our data because we lack measurements of the albedos.

\begin{figure}
\epsscale{1.00}
\begin{center}
\plotone{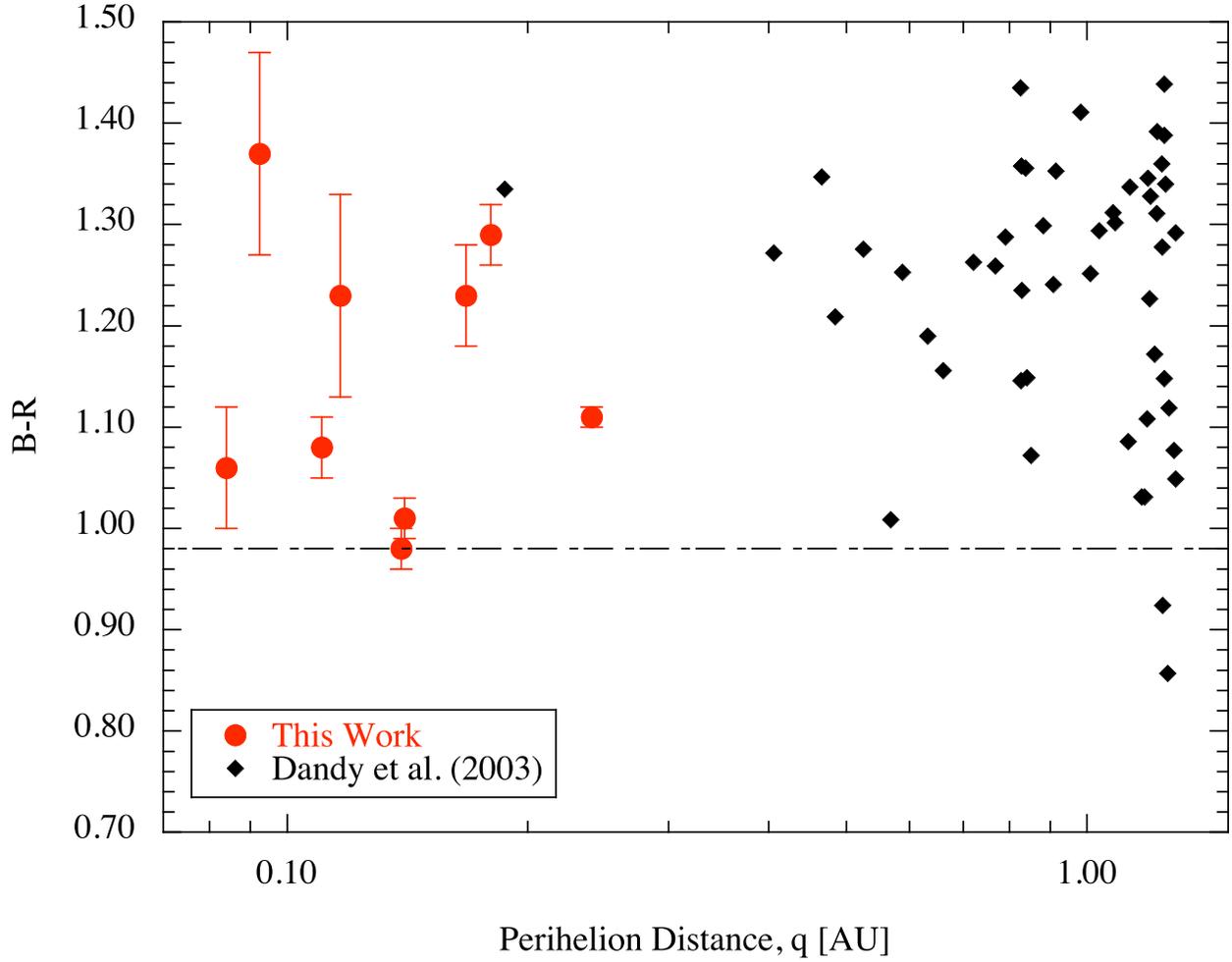}
\caption{Measured colors as a function of perihelion distance.  The horizontal dashed line marks the B-R color of the Sun.   \label{BR_vs_q}
} 
\end{center} 
\end{figure}

Figure (\ref{BR_vs_q}) shows the B-R color index as a function of perihelion distance, $q$.  No dependence of optical color on $q$ is observed in the NSAs.  The Figure also shows the B-R colors of a set of near-Earth objects taken from Table 5 of Dandy et al.~(2003).  These objects typically have $q \sim$ 1 AU, providing a longer distance baseline which should reveal a color-distance trend more clearly, if one were present.  Again, the data provide no visual evidence for any trend.  The Pearson linear correlation coefficient computed from the combined data from Table (\ref{photometry}) and Dandy et al.~(2003) is 0.163 ($N$ = 63). The two-tailed probability of finding this or a larger correlation coefficient from uncorrelated data is $1-P$ = 0.21.  There is no statistical evidence for a color-perihelion distance trend.  Similarly, we find no significant correlation between B-R and the orbital properties $a$, $e$, $i$ or with $H_R$.

\begin{figure}
\epsscale{1.00}
\begin{center}
\plotone{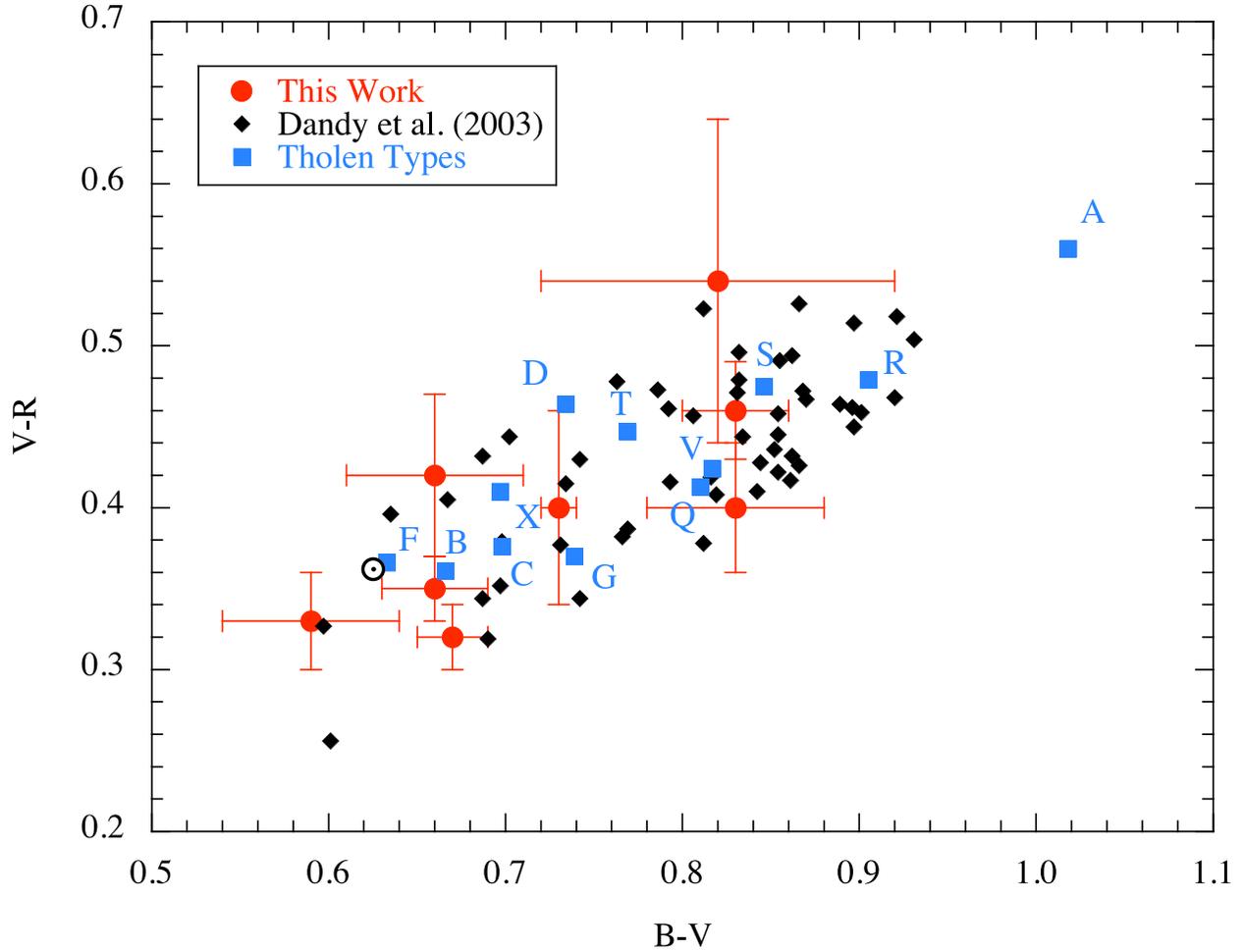}
\caption{Color-color plot showing (red circles) objects from Table (\ref{photometry}) of this paper, (black diamonds) photometry of near-Earth objects from Dandy et al. (2003) and (blue squares) the colors of asteroids falling into the letter-classification scheme of Tholen (1984), as tabulated by Dandy et al. (2003).  The color of the Sun is also marked.   \label{colorcolor}
} 
\end{center} 
\end{figure}

Figure (\ref{colorcolor}) shows the B-V vs.~V-R color plane, where we plot data from Table (\ref{photometry})  with their 1$\sigma$ error bars.  In addition, we show the colors of other near-Earth objects, again from Dandy et al.~(2003).  There is broad overlap between the colors of the NSAs and those of the other near-Earth objects and no evidence for a systematic difference between these two groups.  The planet-crossing asteroids as a whole show B-V and V-R colors consistent with the major asteroid spectral types as defined by Tholen (1984), reflecting their origin in the main belt.

\begin{figure}
\epsscale{1.00}
\begin{center}
\plotone{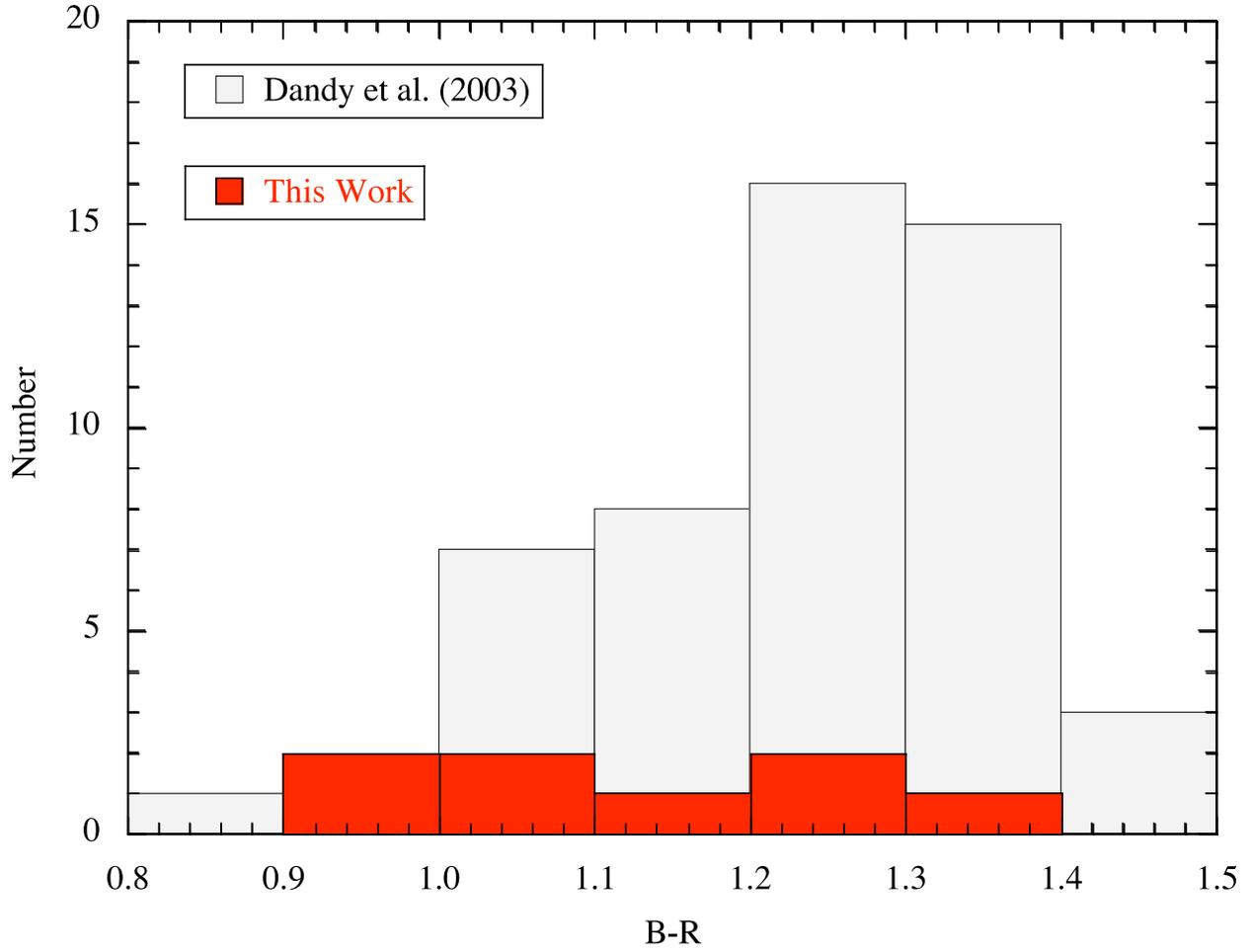}
\caption{Histograms of the B-R color for asteroids in Table (\ref{photometry}) (red) and for near-earth objects from Dandy et al. (2003) (shaded grey).   \label{histogram}
} 
\end{center} 
\end{figure}

To test this more formally, we estimated the likelihood, $P$, that the color histograms for the two groups of asteroids (Figure \ref{histogram}) could be drawn from a single parent population using the Kolmogorov-Smirnov non-parametric test.  The result, $1-P$ = 0.974, is less than a 3$\sigma$ significance (for which $1-P$ = 0.997) and supports the visual impression that no significant color difference exists between the NSAs studied here and the other near-Earth objects discussed by Dandy et al.~(2003).  Observations of a larger sample of NSAs are needed to further address this issue.

\subsection{Coma Search}
We searched visually, without success, for evidence of on-going loss of mass in the form of extended emission around each NSA.  Detection of near-nucleus dust is rendered difficult by the unknown morphology of the coma or tail produced by mass loss.  For example, images of classical comets show a wide range of morphologies ranging from circularly symmetric to highly asymmetric or even linear (Rahe et al.~1969). These differences reflect primarily a competition between the angular distribution and velocity of the ejected dust (tending to produce an extended, roughly symmetric coma) and radiation pressure (tending to deflect particles to one side of the nucleus, producing a linear tail).  The balance of this competition is principally a function of $v_d^2/\beta$, where $v_d$ is the dust ejection speed and $\beta$ is the ratio of the acceleration induced by radiation pressure to that produced by solar gravity.  Parameter $\beta$ is inversely related to particle size, $\overline{a}$.  In sublimation-driven comets,  gas drag relates $v_d$ to $\overline{a}$ in a relatively well-defined way, although the abundance of particles of different sizes remains as an unknown parameter in coma models.   In the objects of the present study, there are several possible mechanisms of ejection (Jewitt 2012) so that a $v_d$ vs.~$\overline{a}$ relation cannot be assumed.  Hence, we lack a simple basis for predicting the morphology to be expected should the NSAs eject dust.

With this as background, we use a simple method to set limits on the presence of coma.  Photometry within concentric projected apertures is used to place  limits to the presence of coma and then to mass loss through a model.  The simplest approximation is that mass loss occurs in steady state and that radiation pressure effects are negligible, in which case the surface brightness of a coma should vary inversely with the angular distance from the nucleus.  Observations are used to set a limit to the surface brightness, $m_{SB}(\theta)$ [magnitudes (arcsec)$^{-2}$] at angular distance, $\theta$ [arcsec]. Then, the magnitude of the steady-state coma encircled inside angle $\theta$ is given by Jewitt and Danielson (1984) as

\begin{equation}
m_{c} = m_{SB}(\theta) - 2.5 \log_{10}(2\pi \theta^2).
\label{jd84}
\end{equation}

\noindent Typically, we set a limit to $m_{SB} \sim$ 26 mag.~(arcsec)$^{-2}$ within an annulus extending from $\theta$ = 3.3\arcsec~to 6.6\arcsec.
Then, the fraction of the scattering cross-section that could be contained within a steady-state coma but have escaped detection is computed from

\begin{equation}
f_c = 10^{[0.4(m_R - m_c)]}
\label{fraction}
\end{equation}

\noindent  provided $m_R < m_c$ and $f_c$ = 1, otherwise.   Here, $m_R$ is the total magnitude as listed in Table (\ref{photometry}) and $m_c$ is from Equation (\ref{jd84}).  

An upper limit to the cross-section of  the coma is calculated from 

\begin{equation}
C_c = 2.24\times10^{22} \pi f_c p_{R}^{-1} 10^{0.4(m_{\odot}(R) - H_{R})} 
\label{geometric}
\end{equation}

\noindent in which $m_{\odot}(R)$ = -27.11 is the apparent red magnitude of the Sun (Drilling and Landolt 2000), $p_{R}$ is the geometric albedo, which we assume to be 0.15, and $f_c$ is from Equation (\ref{fraction}). 

In a distribution of spheres, the combined dust mass, $M_d$, and total cross-section, $C_c$, are related by 

\begin{equation}
M_d = (4/3)\rho \overline{a} C_c
\label{mass}
\end{equation}

\noindent where $\rho$ is the material density and $\overline{a}$ is the average particle radius.  We assume $\rho$ = 3000 kg m$^{-3}$ for all objects. Equations  (\ref{geometric}) and (\ref{mass}) together convert an observational limit on the coma surface brightness  into a dust mass constraint.  

The last step needed to estimate dust mass loss rates is to assign a time-of-residence for the dust particles within the sky annulus.  If the particles are ejected at velocity $v_d$, then the time to cross an annulus of angular width $\delta \theta\arcsec$ is 

\begin{equation}
\tau = k \delta \theta \Delta_{AU} / v_d
\label{time}
\end{equation}

\noindent  where $\Delta_{AU}$ is the geocentric distance expressed in AU and $k$ = 7.3$\times$10$^5$  is a constant equal to the number of meters in 1 arcsec at 1 AU.  We used an annulus of thickness $\delta \theta$ = 3.3\arcsec.

\textit{For purposes of comparison only}, we assume nominal values $\overline{a}$ = 1 $\mu$m and $v_d$ = 1 km s$^{-1}$ in Equations (\ref{mass}) and (\ref{time}), respectively. These values are broadly consistent with the sizes and speeds of the small grains that dominate the optical appearances of most sublimation-driven comets.  The resulting  crossing-times, $\tau$, masses, $M_d$, and mass loss rates, $dm/dt = M_d/\tau$, are listed in Table (\ref{coma}).  By Equations (\ref{mass}) and (\ref{time}) the derived mass loss rates scale in proportion to $\overline{a} v_d$, the product  of the particle size with the ejection velocity.  Millimeter-sized particles ejected at 1 m s$^{-1}$ speeds, like those found in the debris sheets and trails of some comets and active asteroids, would give the same derived mass loss rates as obtained in the nominal case.  Unfortunately, we do not know either $v_d$ or $\overline{a}$ in the NSAs.

Given the nominal dust size and ejection speed, Table (\ref{coma}) shows that existing photometry sets limits on the dust production in the range $\le$0.1 to $\le$1 kg s$^{-1}$.  These are $\le$10$^{-3}$ to $\le$10$^{-4}$ times the mass loss rates shown by active Jupiter family comets when near $R_{au}$ = 1, giving a measure of the generally low level of activity allowed by our imaging data.  Once again, however, the values of $dm/dt$ in Table (\ref{coma}) should not be taken literally, since they are based on arbitrary assumptions about the properties, especially the effective radii and the speeds, of the grains.

Furthermore, the limits to mass loss in Table (\ref{coma}) are based on observations taken when the asteroids were at $R_{au} >$ 1 AU.  They are of little relevance in assessing activity which might occur in the NSAs when closer to the Sun, near perihelion.  This is shown most clearly by the case of (3200) Phaethon, which has never shown evidence for coma when observed against dark sky but in which anomalous perihelion brightening has been associated with the expulsion of dust (Jewitt and Li 2010, Li and Jewitt 2013).  Observations with Keck and other large telescopes are simply not possible at the small heliocentric distances and solar elongation angles of the NSAs when at perihelion.  The solar telescope used to observe Phaethon at perihelion lacks the sensitivity to detect the other NSAs in our sample.

\section{Discussion}

Recent experiments with heated samples of carbonaceous chondrite meteorites provide an interesting context for understanding the present observations (Cloutis et al.~2012).  At low temperatures ($\sim$300 K), these meteorites include abundant phyllosilicates showing characteristic 0.7 $\mu$m absorption bands.  With rising temperature,  progressive dehydration leads to the weakening and eventual disappearance of the 0.7 $\mu$m band.  The spectral slope varies not only with the  temperature, but also depends on the particle size and the nature and spatial distribution of opaque (absorbing) phases in the carbonaceous chondrite.    In general, no simple relation between spectral slope and peak temperature or degree of water loss can be established from laboratory data.  For example, Cloutis et al.~(2012) reported that heated samples of Murchison reddened as the temperature increased to 670 to 770 K, then became bluer at higher temperatures, while a different chondrite (Ivuna) showed no systematic trend over the same temperature range.  These differences reflect the wide range of physical and compositional characteristics of the meteorites, and presumably mirror a wide range of types in the near-Sun asteroid population (c.f.~Figure \ref{colorcolor}).   The duration of heating may also be important, especially given the considerable difference between the timescales used in laboratory experiments (hrs) and those relevant in nature (Myr).

At the highest temperatures (1170 to 1270 K) Cloutis et al.~report that the spectral gradients tend to become neutral to blue and the reflectivity increases to $\sim$10\%.  These trends result, in part, from molecular changes in organics and from agglomeration of metals within the meteorites.  Five of the asteroids in our sample (see Table \ref{orbits}) can attain sub-solar temperatures $>$1000 K when at perihelion, but only two (2002 PD43 and 3200 Phaethon) have optical colors consistent with being neutral or blue.  The others are distinctly redder than sunlight. No spectra of 2002 PD43 are available.  Asteroid Phaethon is a B-type in whose spectrum there is no hint of the 0.7 $\mu$m feature.  While its composition is unknown, at least some B-types have been shown to contain hydrated minerals (Clark et al.~2010, Yang and Jewitt 2010) in which thermal modification would not be surprising (e.g. Montmorillonite, a suggested component of Phaethon (Licandro et al.~2007), thermally decomposes at $\sim$1000 K (Archer et al.~2011)).  Licandro et al.~(2007) have compared the reflection spectrum of Phaethon with heated CI and CM carbonaceous chondrites.  Except for this trend towards blue slopes at the highest temperatures, the laboratory heating experiments show that there is no overall relation between the optical spectral gradients and the past temperature history.  Heated CI and CM chondrites show a range of colors from blue to red, just as do the small-perihelion asteroids in Figure (\ref{BR_vs_q}).

\section{Summary}

We report observations of nine planet-crossing objects selected to have perihelion distances smaller than 0.25 AU.  These objects were observed in order to search for evidence of modification by the high temperatures experienced near perihelion.   We find that

\begin{enumerate}

\item The broadband optical colors are not related to the perihelion distance or other orbital parameters. Neither are they different, statistically, from the colors of  planet-crossing asteroids with larger perihelia.

\item Examination of the near-nucleus region for the presence of dust gives no evidence for on-going mass loss.  Model dependent limits to the mass loss rates in dust fall in the range $\le$0.1 to $\le$1 kg s$^{-1}$. 

\item Combined, our observations provide no evidence to suggest that the physical properties of these small-perihelion objects are measurably influenced by the heat of the Sun.

\end{enumerate}

\acknowledgments
We thank Luca Ricci (LRIS) and Julie Renaud-Kim (Keck) for assistance, Jing Li, Pedro Lacerda and the anonymous referee for comments. This work was supported by a grant to DCJ from NASA's Planetary Astronomy program.

\clearpage

\clearpage

\begin{deluxetable}{llllcccc}
\tablecaption{Orbital Elements 
\label{orbits}}
\tablewidth{0pt}
\tablehead{
\colhead{Object}    & \colhead{$q$\tablenotemark{a}} & \colhead{$a$\tablenotemark{b}} & \colhead{$e$ \tablenotemark{c}}  
& \colhead{$i$\tablenotemark{d}} & \colhead{$T_{BB}(q)$\tablenotemark{e}} & \colhead{$T_{SS}(q)$\tablenotemark{f}} }
\startdata
2004 UL & 0.093 & 1.266 & 0.927 & 23.7  & 912 & 1289\\
2002 PD43 & 0.111 & 2.511 & 0.956 & 26.2 & 834 & 1179 \\
(276033) 2002 AJ129 & 0.117 & 1.371 & 0.915 & 15.5  & 813 & 1150 \\
2006 TC & 0.136 & 1.749 & 0.952 &  27.6  & 754 & 1066 \\
(3200) Phaethon & 0.140 & 1.271 & 0.890 & 22.2   & 743 & 1051 \\
(155140) 2005 UD & 0.163 &  1.275 & 0.872 & 28.8  & 689 & 974  \\
(105140) 2000 NL10 & 0.167 & 0.914 & 0.817 & 32.5  & 680 & 962 \\
(141851) 2002 PM6 & 0.180 & 1.198 & 0.850 & 19.2  & 655 & 926 \\
(225416) 1999 YC & 0.240 & 1.422 & 0.831 & 38.2  & 567 & 802 \\
\enddata


\tablenotetext{a}{ Perihelion distance, AU}
\tablenotetext{b}{ Orbital semimajor axis, AU}
\tablenotetext{c}{ Orbital eccentricity}
\tablenotetext{d}{ Orbital inclination, degree}
\tablenotetext{e}{ Equilibrium isothermal, spherical blackbody temperature at the perihelion distance}
\tablenotetext{e}{ Equilibrium sub-solar blackbody temperature at the perihelion distance}

\end{deluxetable}

\clearpage

\begin{deluxetable}{llclcccc}
\tablecaption{Observational Geometry 
\label{geometry}}
\tablewidth{0pt}
\tablehead{
\colhead{Object}  & \colhead{UT Date}  & \colhead{$q$\tablenotemark{a}} & \colhead{$R$\tablenotemark{b}} & \colhead{$\Delta$\tablenotemark{c}}  
& \colhead{$\alpha$\tablenotemark{d}} }
\startdata
2004 UL &2012 Oct 13 & 0.093 & 2.429 & 1.496 & 10.6  \\

2002 PD43  & 2010 Aug 10 &0.111 & 1.671 & 0.665 &   6.8 \\
                       
(276033) 2002 AJ129 & 2009 Mar 30 & 0.117 & 2.609 & 1.717 &  12.2 \\

2006 TC & 2010 Aug 10 & 0.136 & 1.363 & 1.009 &  47.8 \\
                 & 2010 Sep 10 & ... & 1.748 & 0.952 & 28.0  \\

(3200) Phaethon     & 2010 Sep 10 & 0.140  & 1.510 & 1.709 &  35.8 \\

 & 2012 Oct 14 & ... & 2.184 & 1.345 & 18.1  \\

(155140) 2005 UD & 2005 Nov 21-22 & 0.163 & 1.592 & 0.960 & 35.8 \\
 & 2007 Oct 12 &... & 2.601 & 1.819 &  16.4  \\
 
 (105140) 2000 NL10 & 2012 Oct 13 & 0.167 & 1.574 & 1.326 & 38.9 \\
 & 2012 Oct 14 & ... & 1.578 & 1.339 & 39.1 \\
 
(141851) 2002 PM6 & 2012 Oct 13 & 0.180 & 1.284 & 0.466 & 43.2 \\
(225416) 1999 YC & 2007 Oct 04 & 0.240 & 2.603 & 1.912 & 18.7 & \\
 & 2007 Oct 12 & ...  & 2.601& 1.819 & 16.4 & \\ \\
\enddata


\tablenotetext{a}{Perihelion distance, AU}
\tablenotetext{b}{Heliocenric distance, AU}
\tablenotetext{c}{Geocentric distance, AU}
\tablenotetext{d}{Phase angle, degree}

\end{deluxetable}
\clearpage

\begin{deluxetable}{llccccccc}
\tabletypesize{\scriptsize}
\tablecaption{Photometry
\label{photometry}}
\tablewidth{0pt}
\tablehead{
\colhead{Object} & \colhead{UT Date}  & \colhead{$m_R$ \tablenotemark{a}}  & \colhead{$B-V$} & \colhead{$V-R$} & \colhead{$B-R$}& \colhead{$m_R(1,1,\alpha)$\tablenotemark{b}} & \colhead{$H_R$  \tablenotemark{c} }}
\startdata
2004 UL & 2012 Oct 13 & 22.24$\pm$0.07 & 0.82$\pm$0.10 & 0.54$\pm$0.10 & 1.37$\pm$0.10 & 19.44 & 18.77 \\
2002 PD43 & 2010 Aug 10 & 19.63$\pm$0.02 & 0.66$\pm$0.05 & 0.42$\pm$0.05 & 1.08$\pm$0.03 & 19.40 & 18.89 \\
(276033) 2002 AJ129& 2009 Mar 30 & 22.36$\pm$0.05 & -- & -- & 1.23$\pm$0.10 & 19.10 & 18.37 \\
2006 TC & 2010 Aug 10 & 20.44$\pm$0.05 & 0.59$\pm$0.07 & -- & 1.26$\pm$0.10 & 19.75 & 17.95 \\
 & 2010 Sep 10 & 20.66$\pm$0.02 & 0.60$\pm$0.08 & 0.33$\pm$0.03 & 0.93$\pm$0.08 & 19.56 & 18.31 \\
(3200) Phaethon & 2010 Sep 10 & 17.52$\pm$0.01 & 0.67$\pm$0.02 & 0.32$\pm$0.02 & 0.99$\pm$0.02 &15.46 & 14.00\\
 & 2012 Oct 14 & 17.25$\pm$0.05 &-- & -- & 0.85$\pm$0.07 & 14.91 & 13.97  \\

2005 UD\tablenotemark{d} & 2007 Nov 21-22 & 19.45$\pm$0.02 & 0.66$\pm$0.03 & 0.35$\pm$0.02 & 1.01$\pm$0.02 & 18.53 & 17.08\\
105140 2000 NL10  & 2012 Oct 13 & 18.19$\pm$0.05 & -- & -- & -- & 16.59 & 15.04 \\
  & 2012 Oct  14 & 18.84$\pm$0.04 & 0.83$\pm$0.05 & 0.40$\pm$0.04 & 1.23$\pm$0.05 & 17.22 & 15.66\\
(141851) 2002 PM6 & 2012 Oct 13 & 18.00$\pm$0.02 & 0.83$\pm$0.03 & 0.46$\pm$0.03 & 1.29$\pm$0.03 &19.12 & 17.45\\
1999 YC\tablenotemark{e} & 2007 Oct 04 & 21.14$\pm$0.01 & -- & 0.40$\pm$0.06 & 1.07$\pm$0.07 & 17.55 & 16.70\\
 & 2007 Oct 12 & 21.03$\pm$0.02 & 0.73$\pm$0.01 & -- & 1.11$\pm$0.01 & 18.16 & 16.78\\

Solar Color & -- & -- & 0.63 & 0.36 & 0.99 & -- & --  \\



\enddata


\tablenotetext{a}{Apparent red magnitude}
\tablenotetext{b}{Magnitude at unit heliocentric and geocentric distances and the observed phase angle, $\alpha$ (see Table (\ref{geometry}))}
\tablenotetext{c}{Absolute red magnitude computed from Equation (\ref{inverse}) using the HG formalism with $g$ = 0.15.}
\tablenotetext{d}{Observations from Jewitt and Hsieh 2006}
\tablenotetext{e}{Observations from Kasuga and Jewitt 2008}

\end{deluxetable}

\clearpage

\begin{deluxetable}{llcccccc}
\tabletypesize{\scriptsize}
\tablecaption{Coma Constraints
\label{coma}}
\tablewidth{0pt}
\tablehead{
\colhead{Object} &\colhead{UT Date}    & \colhead{$f_c$\tablenotemark{a}}  & \colhead{$C_c$\tablenotemark{b}} & \colhead{$M_d$\tablenotemark{c}} & \colhead{$\tau_c$\tablenotemark{d}}& \colhead{$M_d/\tau_c$\tablenotemark{e}} }
\startdata
2004 UL & 2012 Oct 13 &   1.00  & 0.21 & 834 & 3600 & 0.2 \\
2002 PD43 & 2010 Aug 10 &    0.20 & 0.04 & 149 & 1600 & 0.1 \\
(276033) 2002 AJ129& 2009 Mar 30 &    1.00 & 0.30 & 1206 & 4136 & 0.3 \\
2006 TC & 2010 Aug 10 &    0.44 & 0.20  & 781  & 2400 & 0.3 \\
 & 2010 Sep 10 &    0.51 & 0.16 & 650 & 2288 & 0.3 \\
(3200) Phaethon & 2010 Sep 10 &    0.03 & 0.51 & 2025 & 4120 & 0.5 \\
 & 2012 Oct 14 &    0.02 & 0.35 & 1388 & 3240 &  0.4 \\

2005 UD & 2007 Nov 21-22   &  0.17 & 0.17 & 672 & 2312 & 0.3 \\
105140 2000 NL10  & 2012 Oct 13   & 0.15 & 0.97 & 3885 & 3194 & 1.2\\
  & 2012 Oct  14   &  0.09 & 0.33 & 1317 & 3225 & 0.4\\
(141851) 2002 PM6 & 2012 Oct 13   &  0.04 & 0.03 & 112 & 1122 & 0.1\\
1999 YC & 2007 Oct 04   &  0.79 & 1.11& 4436 & 4606 & 1.0 \\
 & 2007 Oct 12   &  0.71 & 0.93 & 3700 & 4382 & 0.8 \\

\enddata


\tablenotetext{a}{Fraction of the cross-section which could be contributed by a steady-state coma, from Equation (\ref{fraction})}
\tablenotetext{b}{Cross-section of coma (km$^2$), from Equation (\ref{geometric})}
\tablenotetext{c}{Mass of coma (kg) inside angular radius $\theta$ = 3.3\arcsec, from Equation (\ref{mass})}
\tablenotetext{d}{Annulus crossing-time (s), from Equation (\ref{time}) }
\tablenotetext{e}{Derived mass loss rate in dust (kg s$^{-1}$), from Equations (\ref{mass}) and (\ref{time})}

\end{deluxetable}

\clearpage

\clearpage

\clearpage

\clearpage

\end{document}